\documentclass[doublecol]{epl2}
\usepackage{graphicx,amsmath,amssymb,mathrsfs}
\usepackage{epsfig}

\begin{document}

\title{Multifractal information production of the human genome}

\author{Christian Beck\inst{1} and Astero Provata\inst{1,2}  }

\institute{
\inst{1}
Queen Mary University of London, School of Mathematical Sciences, Mile End Road, London E1 4NS, UK\\
\inst{2}
Institute of Physical Chemistry, 
National Center for Scientific Research "Demokritos",
15310 Athens, Greece
}

 {Structure and Organisation in Complex Systems}
 {Fractals}
 {Biomolecules:Theory, Modelling and Computer Simulations}

\abstract{
We determine the Renyi entropies $K_q$ of symbol sequences generated by human chromosomes.
These exhibit nontrivial behaviour as a function of the scanning parameter $q$. 
In the thermodynamic
formalism, there are phase transition-like phenomena 
close to the $q=1$ region. We develop a
theoretical model for this based on the superposition of two multifractal sets, which can be
associated with the different statistical properties
of coding and non-coding DNA sequences. This model is in good agreement with
the human chromosome data.
}

\maketitle

DNA symbol sequences exhibit a very complicated dynamical 
 structure. There are long-range
correlations \cite{li:1992,peng:1992,voss:1992,ebeling:1992,mantegna:1994,scaling,arneodo:1998,arneodo:1998-2,usatenko:2003,afreixo:2004,carpena:2007} 
which are particularly strong for the non-coding sequences (DNA sequences which do not code
for the production of proteins)
whereas the coding sequences demonstrate characteristics similar
to random-like 
processes \cite{li:1992,peng:1992,mantegna:1994,scaling}. The way in which coding and
non-coding sequences alternate in the DNA of many organisms 
is described by a multifractal \cite{yu:2001,gutierrez:2001,su:2009,provata:2010}. 
Various approaches have been suggested to map DNA sequences onto 
the dynamics of
an associated dynamical system, such as correlated random walks \cite{peng:1992,scaling}, or
to provide a suitable measure representation by formally mapping
DNA sequences onto points of the unit interval\cite{gutierrez:2001,hao:2000}. 
The associated measures,
investigated in detail by Yu et al. for a large variety of organisms \cite{yu:2001},
exhibit a non-trivial spectrum of Renyi dimensions.

In this paper we directly apply the known
symbolic dynamics techniques of the thermodynamic
formalism of dynamical systems\cite{beck:book1995,tel, szep} to DNA symbol sequences. 
For our data analysis we will concentrate mostly
on the human genome (chromosome 10) as a working example. For DNA the symbol space
contains 4 different symbols A,G,T,C denoting the four nucleotides
(Adenine, Guanine, Thymine and Cytosine). 
Translations along the DNA string can be regarded as
a shift of (correlated) symbols.
We are interested in the average information production produced by this shift,
and in the set of all higher-order correlations of the symbols.
This can be measured by various quantities which weight the rare and frequent
symbol sequences in a different way.
In dynamical systems theory, for a system with a generating partition,
one defines the dynamical Renyi entropies as
\begin{eqnarray}
K_q=\lim_{N\to \infty} \frac{1}{N} \frac{1}{1-q} \ln \sum_{i_1, \ldots , i_N}
 p(i_1,i_2, \ldots ,i_N)^q, \>\> q\ne 1 \\
\nonumber
K_1=\lim_{N\to \infty} \frac{1}{N} \sum_{i_1, \ldots , i_N} 
p(i_1,i_2, \ldots ,i_N)\ln p(i_1,i_2, \ldots ,i_N) 
\label{eq01}
\end{eqnarray}
Here $p(i_1, i_2, \ldots , i_N)$ denotes the probability of the symbol sequence
$i_1, i_2, \ldots , i_N$. $N$ denotes the length of the sequence and
 $q$ is a parameter taking real values.
The above sum is taken over all allowed symbol sequences $i_1, i_2, \ldots ,i_N$,
i.e. over all sequences with $p(i_1, \ldots , i_N)\not= 0$.
$K_1$ is the Kolmogorov-Sinai entropy, a very important invariant in dynamical
system theory. $K_0$ is the topological entropy, which counts the growth
rate of allowed symbol sequences for $N\to \infty$.
A much more complete characterisation is via the set of
all $K_q$ with $q\in (-\infty, \infty)$. 
These quite generally measure the information production
of the dynamical system under consideration.
From this set one can proceed to the spectrum of dynamical crowding
indices by Legendre transformation (see e.g. \cite{beck:book1995,takayasu:book1990} for details).

\begin{figure}
\includegraphics[clip,width=0.45\textwidth,angle=0]{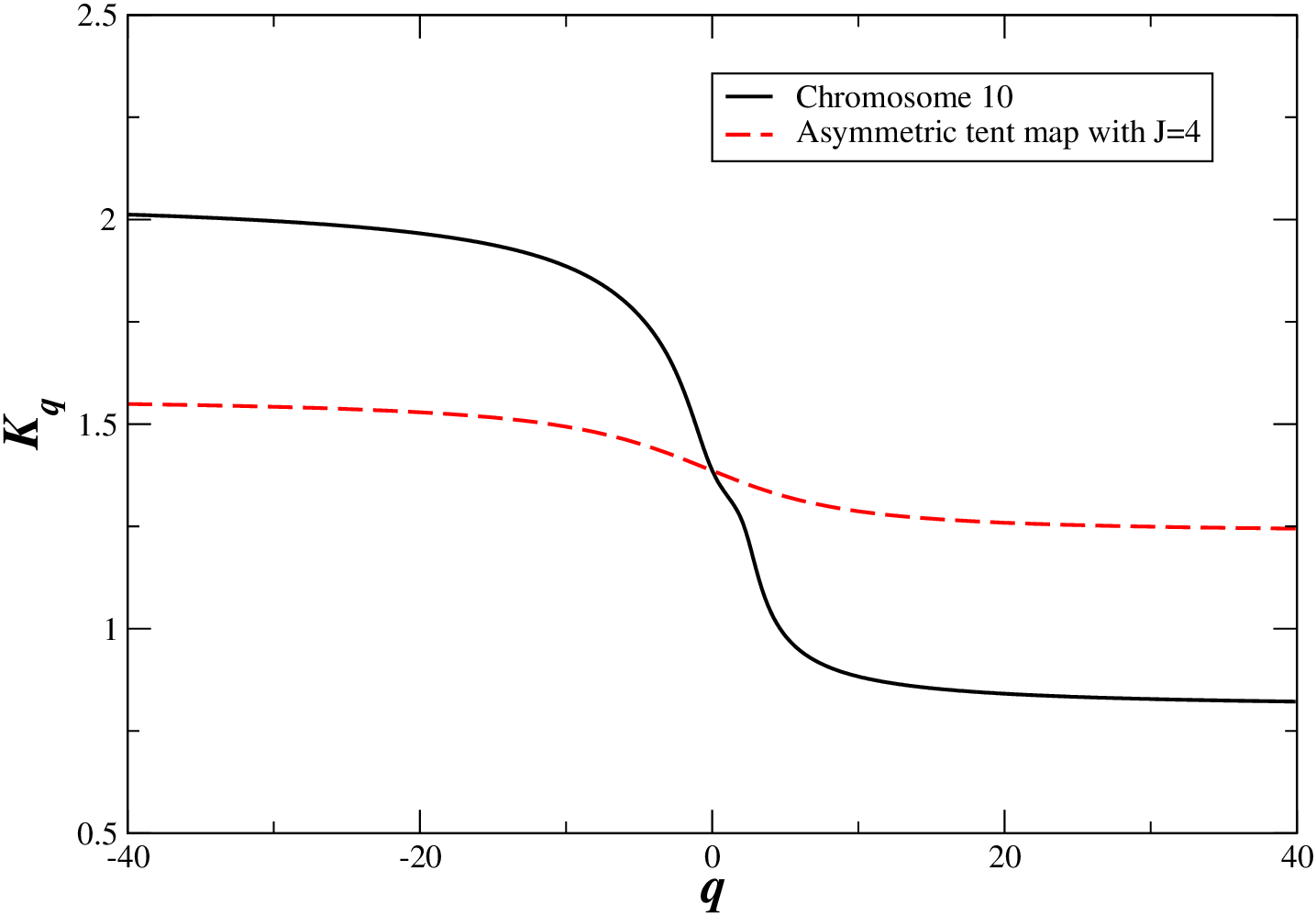} 

\caption{\label{fig:01}{} (Colour online) 
 The spectrum of Renyi entropies $K_q$ for human chromosome 10
(black solid line). The dashed line shows the corresponding spectrum
of an asymmetric tent map with 4 symbols and the same 1-point probabilities
as chromosome 10.} 
\end{figure}

For the standard  Bernoulli shift of $J$ different symbols, the symbols are statistically
independent and occur with equal probability $p=1/J$. We thus obtain
$p(i_1, \ldots , i_N)=p^N=J^{-N}$ and $K_q=\ln J$, independent of $q$.
If there are non-trivial correlations, 
and non-uniform probabilities, as is the case for DNA sequences, then the
spectrum of $K_q$ becomes nontrivial. As an example,
in fig.\ref{fig:01} the solid black line shows the multifractal $K_q$ spectrum
 obtained for the human chromosome 10.
The spectrum was numerically evaluated by taking into
account all symbol sequences up to length $N=8$. This length is adequate for
representing the asymptotic spectrum, which is already reached for values 
$N\geq 6$, as was also reported in references \cite{yu:2001,provata:2010}.

\par
\begin{figure}
\includegraphics[clip,width=0.23\textwidth,angle=0]{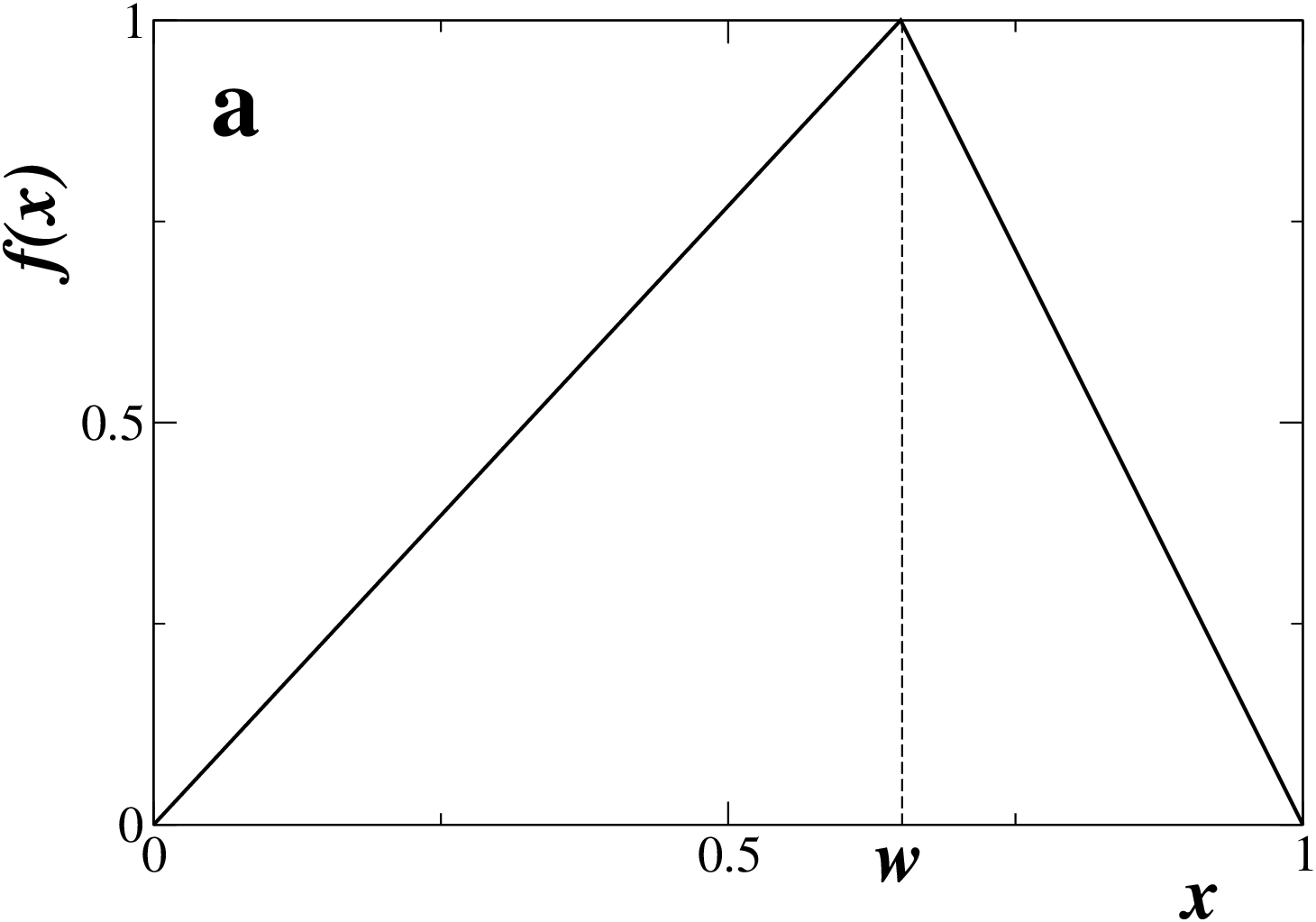}
\includegraphics[clip,width=0.23\textwidth,angle=0]{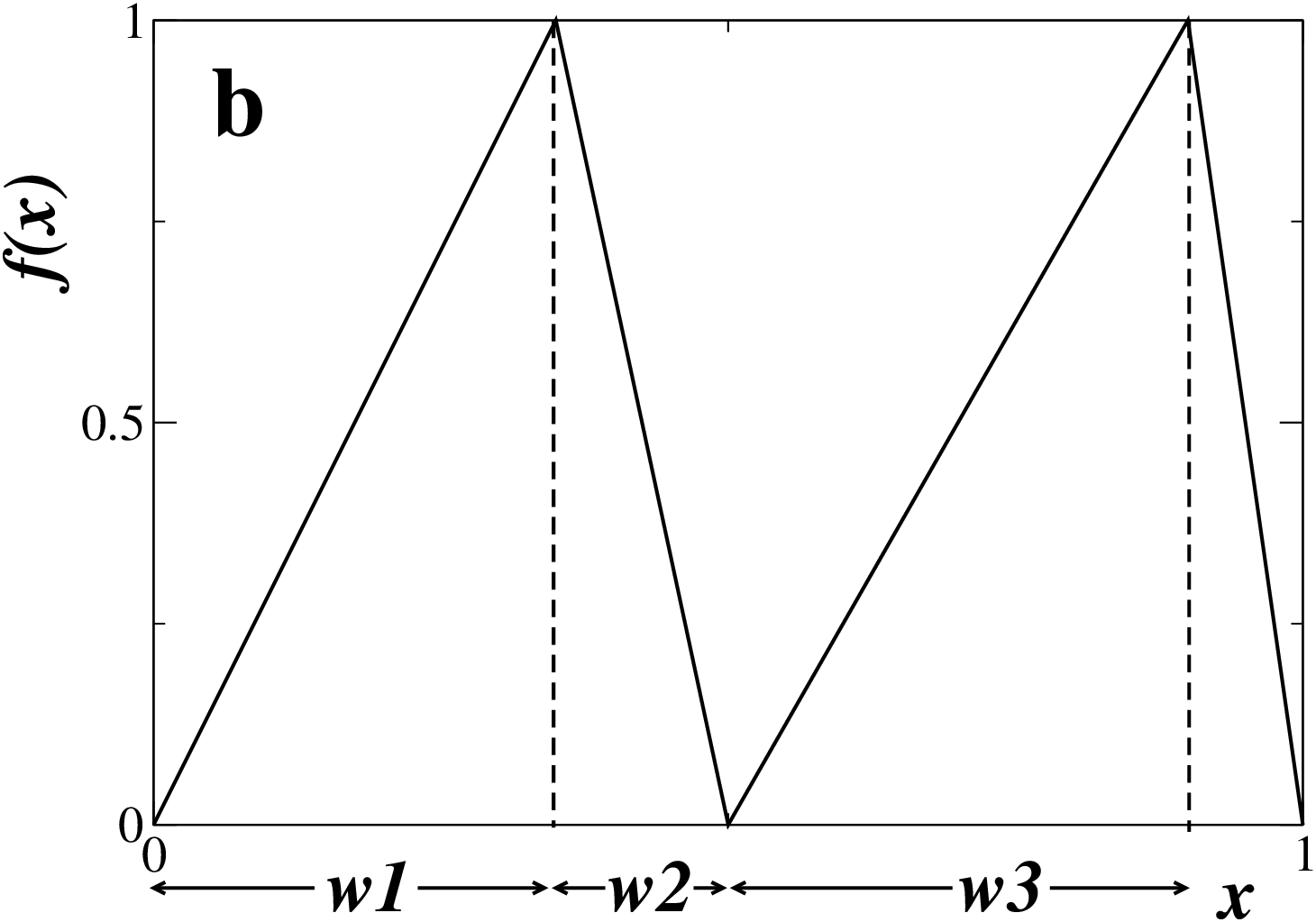}
 

\caption{\label{fig:02}{} (Colour online) 
Example of an asymmetric tent map with (a) 1 maximum (shift of 2 symbols)
and (b) 2 maxima (shift of 4 symbols).} 
\end{figure}

 Our goal is to compare the information production of symbol sequences
of the human genome with those generated by simple examples of chaotic maps.
A simple example of a dynamical system with a nontrivial $K_q$
spectrum is the asymmetric tent map (fig. \ref{fig:02}a), given on the unit interval $[0,1]$
by 
\begin{equation}
f(x)= \left\{
\begin{array}{ll} 
\frac{x}{w} & \mbox{for}
\, 0 \leq x \leq w \\
\frac{1-x}{1-w} & \mbox{for} \, w \leq x \leq 1
\end{array}
\right.
\label{eq02}
\end{equation}

The generating partition for this map corresponds to
the two intervals $I_1=[0,w]$ an $I_2=[w, 1]$. We
may wright the symbol '1' if an iterate $x_n$ of $f$ is in $I_1$
and '2' if it is in $I_2$. The Renyi entropies
for this simple model system are given by
\begin{eqnarray}
K_q&=&\frac{1}{1-q} \ln (w^q +(1-w)^q), \>\> q\ne 1 \\
\nonumber
K_1&=&w\ln w+(1-w)\ln (1-w)
\label{eq03}
\end{eqnarray}
The above chaotic dynamical system generates symbol sequences consisting
of just two different symbols. An obvious generalisation
is to $J$ different symbols, where the 
corresponding piecewise linear map has $J/2$
maxima (fig. \ref{fig:02}b). In this case the $K_q$ are given by
\begin{eqnarray}
K_q &=& \frac{1}{1-q} \ln (w_1^q+w_2^q+ \ldots + w_J^q),\>\> q\ne 1 \\
\nonumber
K_1 &=& \sum_{i=1}^J w_i \ln w_i
\label{eq04}
\end{eqnarray}
with $w_1+w_2+\ldots +w_J=1$. The parameters $w_j$ correspond
to the 1-point probabilities of the occurrences of the symbols $j$.
\par
For human Chromosome 10, the observed values of 1-point
symbol probabilities are
$w_1=w_A=0.291921, w_2=w_C=0.207966,w_3=w_G=0.207859$ and $w_4=w_T=0.292219$
\cite{provata:2010}. 
The entropies $K_q$ of the human genome can neither be fitted by the above
simple model with $J=2$, which in 
the multifractal language corresponds to a two-scale Cantor set
with a multiplicative measure, nor
using $J=4$, which corresponds to
a 4-scale Cantor set, choosing the same 1-point probabilities as
observed.
This is shown in fig. \ref{fig:01}: The $q$-dependence
of the chromosomes data is much more pronounced than that of the
corresponding asymmetric chaotic map that shifts 4 symbols.
We thus need a more sophisticated approach to reproduce the
observed multifractal information production of the human genome.

The idea developed in the sequel
is to take into account the different dynamical properties of
the coding and noncoding strings which constitute the chromosomes.
 The symbol sequence probabilities are, 
in general, different for each of those
regions,
and are denoted by $p^{(c)}(i_1, \ldots . i_N)$ and 
$p^{(nc)}(i_1, \ldots , i_N)$, respectively. 
In the following, inspired by the multifractal formalism,
we consider sequences of size $N$ as part of longer sequences
and we write $N=-\log \epsilon$, where $\epsilon$ is the 
partition 'box size'. The limit $N\to \infty$
corresponds to `box size' $\epsilon \to 0$, and the
$K_q$ are then identical (up to a multiplicative factor)
 to the $D_q$ of a multifractal that
encodes the dynamical properties.

When the dynamical partition function
\begin{equation}
Z(q) :=\sum_{i_1, \ldots , i_N} p(i_1, \ldots , i_N)^q
\sim \epsilon^{(q-1)K_q} \label{eq05}
\end{equation}
is evaluated, there are contributions from both types of strings.
We thus have
\begin{eqnarray}
Z(q) &\approx&  N_c \sum p^{(c)}(i_1, \ldots , i_N)^q +N_{nc} \sum
p^{(nc)}(i_1, \ldots i_N)^q  \nonumber \\
\, & \sim & N_c \epsilon^{(q-1)K_q^{(c)}} + N_{nc} \epsilon^{(q-1)K_q^{(nc)}},
\label{eq06}
\end{eqnarray}
where the numbers $N_c$, $N_{nc}$ determine how many strings are in
the coding and
non-coding region, respectively. If $N_c,N_{nc}$ are independent of $\epsilon$,
then the Renyi entropies of the entire system are given by the
term that dominates the partition function for $\epsilon \to 0$, i.e
\begin{equation}
K_q = \left\{
\begin{array}{ll}
\min (K_q^{(c)}, K_q^{(nc)}) & \mbox{for} \, q>1 \\
\max (K_q^{(c)}, K_q^{(nc)}) & \mbox{for} \, q<1.
\end{array}
\right.
\label{eq07}
\end{equation}
In the thermodynamic formalism of dynamical systems,
this means that the free energy $(q-1)K_q$ exhibits a phase transition (non-analytic
behaviour) at the critical value $q_{critical}=1$ (see also \cite{szep}
for other systems 
exhibiting phase transitions in the Renyi entropies).
Clearly such a behaviour can only be
seen if one uses other entropy measures than the usual KS entropy 
(corresponding to $q=1$) for the investigation
of the information production of the human genome.
This once again illustrates the importance to study the entire
multifractal spectrum $K_q$.

\begin{figure}
\includegraphics[clip,width=0.45\textwidth,angle=0]{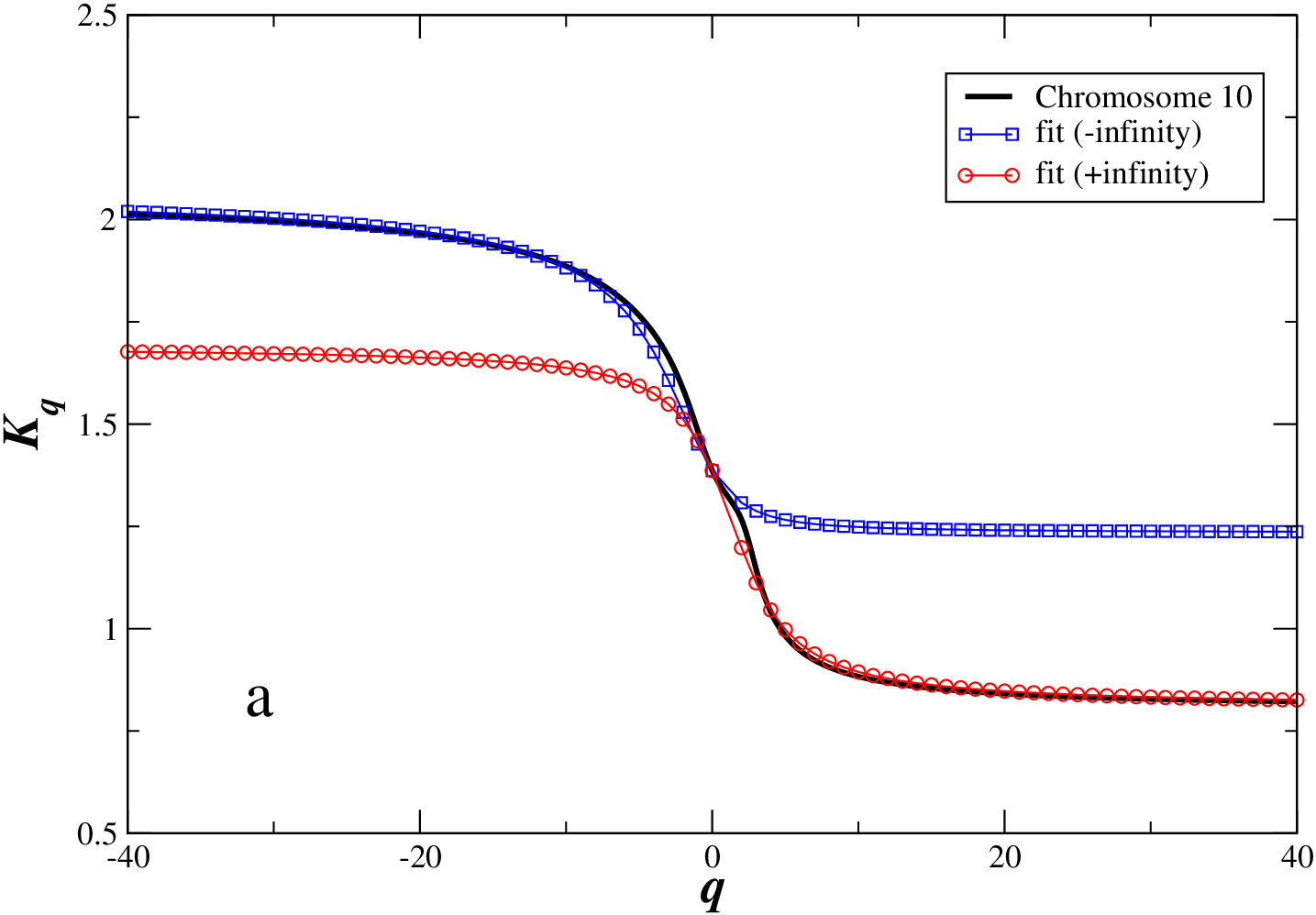}\\
\includegraphics[clip,width=0.45\textwidth,angle=0]{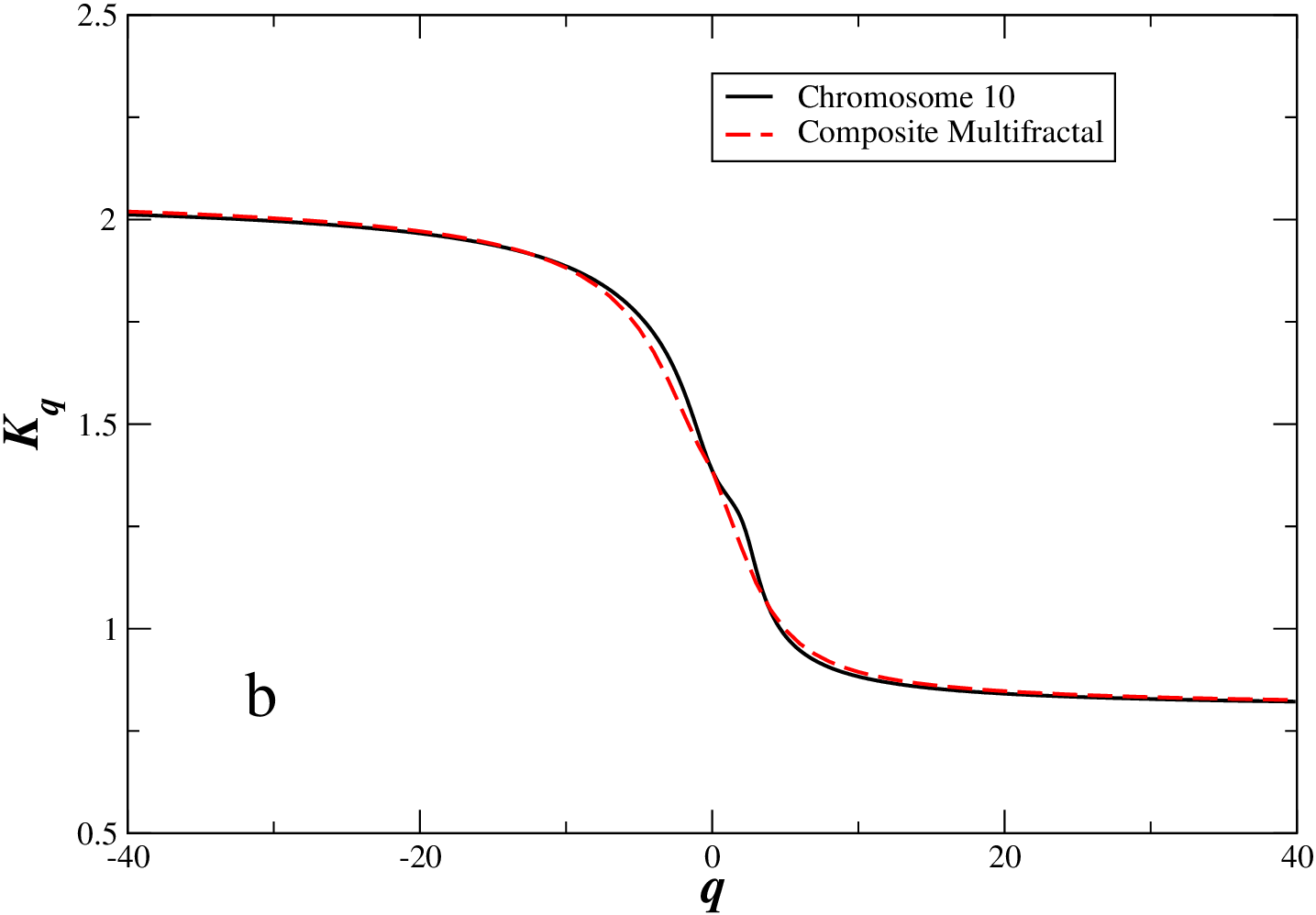} 

\caption{\label{fig:03}{} (Colour online) a) Separate approximations to the
$K_q$ spectrum for $q\to -\infty$ (blue squares) and $q\to +\infty$ 
(red circles).
b) Composite multifractal spectrum (red triangles) and multifractal spectrum of
Chromosome 10, organism Homo Sapiens (solid black line). }
\end{figure}

The above simple phase transition
model of $K_q$ agrees well with the genome data, see fig.
\ref{fig:03}. Figure  \ref{fig:03}a shows two approximations of the human 
chromosome data via
two different multifractal sets.
For the modelling multifractal sets with $J=4$ different symbols were
taken into account, since the genome consists of 4 nucleotides. For simplicity
only one effective scale $w_1$ was introduced into each of the two sets, leading to
\begin{eqnarray}
K_q &=& \frac{1}{1-q} \ln (w_1^q+3 w_2^q),\>\> q\ne 1 \\
\nonumber
K_1 &=& w_1 \ln w_1 + 3 w_2 \ln w_2
\label{eq071}
\end{eqnarray}
where $w_2=(1-w_1)/3$.  
The first one approximates
well the chromosome 10 data when $q\to \infty$ with $ w_1=0.447$ 
but fails in the region
$q\to -\infty$, see fig. \ref{fig:03}a (red circles). 
The second multifractal set approximates the data in the opposite
region, with $w_1=0.126$, see fig. \ref{fig:03}a (blue squares). 
In fig \ref{fig:03}b the red-dashed line is a composite of the 
two multifractal sets, based on forming the maximum, respectively
the minimum, according to eq.~\ref{eq07}. This
approximates the data well in the entire $q$-region.
In fig.\ref{fig:03} the values of the limit  entropies $K_\pm\infty$
were fitted to give the best coincidence with the data. Note that
the region $q\to -\infty$ is dominated by very rare symbol sequences
and the region $q\to +\infty$ by the most frequent ones.
Also, it should be clear that
finite size effects demonstrated in the genomic data make a sharp phase transition
unobservable since, as in our numerical
analysis, only  symbol sequences of finite size 
are investigated. Our hypothesis in the following
is to associate the blue curve (squares) in fig. \ref{fig:03} with
the non-coding sequences and the red curve (circles) with the coding ones.
\par
\begin{figure}
\includegraphics[clip,width=0.45\textwidth,angle=0]{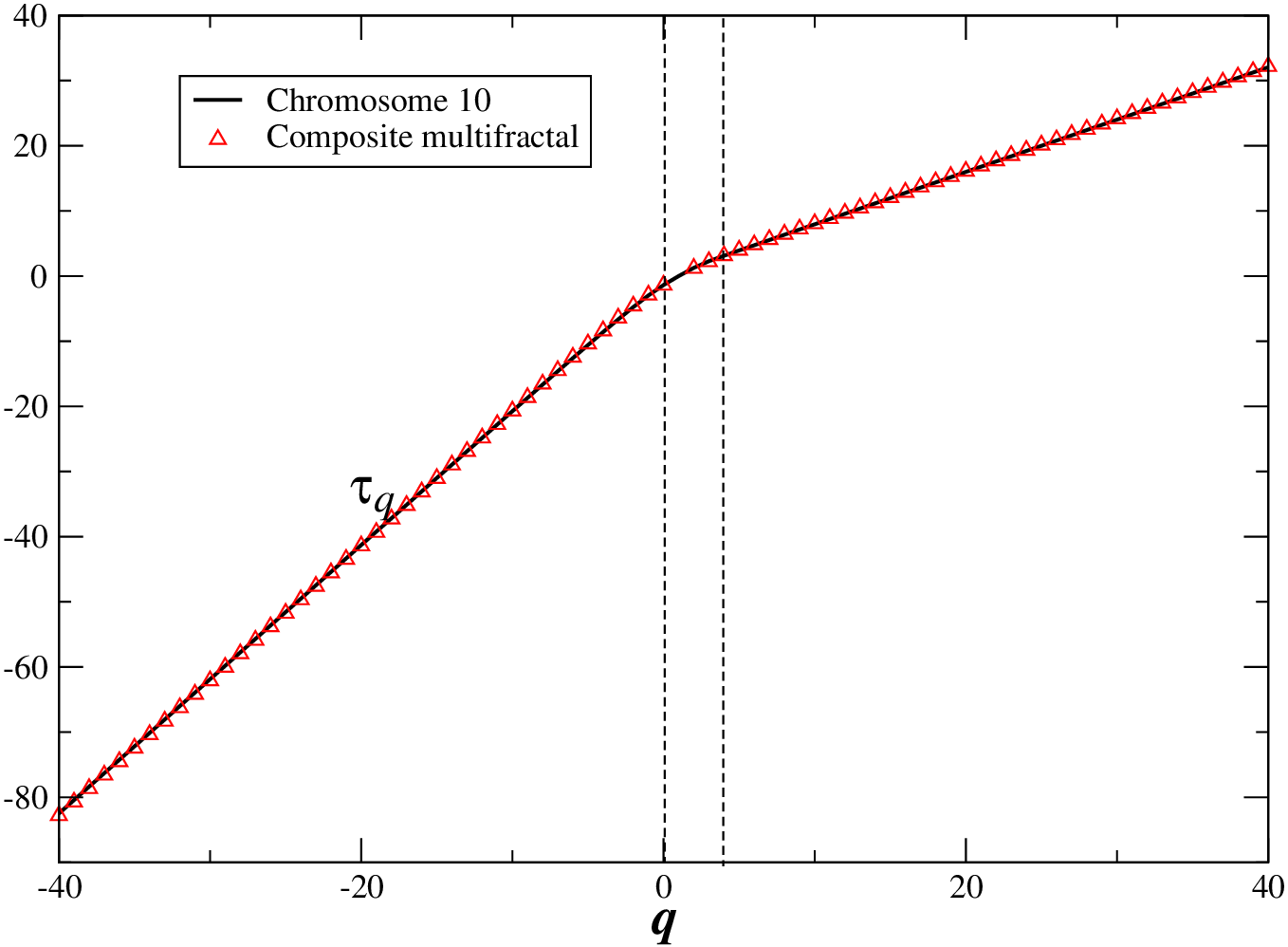}
 

\caption{\label{fig:04}{} (Colour online) 
The $\tau_q$ of  
Chromosome 10, organism Homo Sapiens (solid black line)
and of the
composite multifractal spectrum (red triangles) }
\end{figure}

 
In the thermodynamic formalism of dynamical systems, the
role of the free energy is played by the function $\tau_q=(q-1)K_q$
rather than $K_q$ itself
\cite{beck:book1995}. It is therefore useful to analyze this function
in somewhat more detail. $\tau_q$ is shown in fig.
\ref{fig:04}, with the solid black line representing the human chromosome 10
and the red triangles originating from the composite model. Again we see
evidence for the presence of a critical value $q_{critical}$
with phase-transition-like behaviour.
An abrupt change of slope is clearly observable
in the area $0\le q \le 4$, though of course
the precise value of the
critical $q$-value cannot be located due to finite size effects. 
Our model predicts that $\tau_q$ is a continuous but non-differentiable
function of $q$ at $q_{critical}=1$, which in the thermodynamic
analogy corresponds to a 1st-order phase transition.
The relevant transition area is designated in 
fig. \ref{fig:04} by two perpedicular dashed lines. 

\par So far
our composite multifractal model shows a phase transition
at $q=1$, since by construction the two Cantor sets
were joint at the $q=1$ scale, see eq. \ref{eq07}.
On the other hand, it is known that the numbers $N_c$ and $N_{nc}$ 
can depend on $\epsilon$
in a significant way. Long range correlations
are demonstrated in the noncoding,
while short range ones are displayed by the coding sequences 
\cite{peng:1992,li:1992,scaling,provata:2010}.
The structure of (mostly) noncoding sequences, as intervowen with coding
sequences,
forms a (multi-)fractal as well.
This means the above 
numbers $N_c$ and $N_{nc}$ scale with $\epsilon$ and thus the
critical value $q_{critical}$ 
can shift to different values. This is clearly observed in the 
present data, both in the $K_q$ spectrum (see fig. \ref{fig:03})
and in the $\tau_q$ one (see fig. \ref{fig:04}).
Fig. \ref{fig:04} indicates that
the critical point is slightly displaced to a value 
$q_{critical}\approx 2 >1$. As we shall see below,
this behaviour can be understood from 
the domination of the long range correlated noncoding 
sequences, $N_{nc}>>N_c$, which are known to cover approximately 97\%
of the human genome.

Mathematically, if we assume that the 
coding sequences scale as
\begin{equation}
N_c \sim \epsilon^{-d_c}
\end{equation}
and the non-coding ones as
\begin{equation}
N_{nc} \sim \epsilon^{-d_{nc}},
\end{equation} 
then the critical point $q_{critical}$ is determined by the
relative dominance of the two exponents in eq. \ref{eq06}, i.e.
by the condition
\begin{equation}
(q_{critical}-1)K_q^{(c)}-d_c=(q_{critical}-1)K_q^{(nc)}-d_{nc},
\end{equation}
which, depending on the numbers $d_c$ and $d_{nc}$, can shift the
critical value away from 1. Solving for $q_{critical}$ we obtain
\begin{equation}
q_{critical}=1+\frac{d_{nc}-d_c}{K_q^{(nc)}-K_q^{(c)}}.
\label{qcrit}
\end{equation}
At $q\approx 2$ we see from fig. \ref{fig:03} that
$K_q^{(nc)}$ (blue squares) is bigger than $K_q^{(c)}$ (red circles).
Hence eq. \ref{qcrit}
implies that $d_{nc}>d_c$. This, on the other hand, implies
\begin{equation}
N_{nc} \sim \epsilon^{-d_{nc}} >>N_c \sim \epsilon^{-d_c},
\end{equation}
consistent with the fact that the number $N_{nc}$ of
non-coding sequences dominates
over the number $N_c$ of coding ones.

To conclude, we have shown that the information production
of the human genome, if regarded as a shift of the four symbols
$A,C,G,T$, is very
complex and can only be fully understood by considering the entire spectrum
of Renyi entropies $K_q$. The multifractal structure
can be approximated to a great extent by a superposition of
two processes, one describing the system for $q >q_{critical}
$ and one for $q<q_{critical}$, corresponding
roughly to coding and non-coding DNA characteristics.

\end{document}